# Ultrafast wide-field 3D topography with extended depth of field

**QIANYI WEI[1], JIELEI NI[1,*], YUQUAN ZHANG[1], ZHANGYU ZHOU[1], SHUOSHUO ZHANG[1,2], ZHIYONG TAN[1], JIAHUI PAN[1], XIAOCONG YUAN[1**], AND CHANGJUN MIN[1,***]**

[1]*Nanophotonics Research Center, Institute of Microscale Optoelectronics & State Key Laboratory of Radio Frequency Heterogeneous Integration, Shenzhen University, Shenzhen 518060, China*

[2]*Key Laboratory of Light Field Manipulation and Information Acquisition, Ministry of Industry and Information Technology, School of Physical Science and Technology, Northwestern Polytechnical University, Xi'an 710129, China*

Corresponding author: *nijielei@szu.edu.cn; **xcyuan@szu.edu.cn;***cjmin@szu.edu.cn

**Ultrafast optical imaging has enabled direct observation of femtosecond–nanosecond dynamics, yet three-dimensional (3D) dynamic measurements at high numerical aperture (NA) remain hindered by the intrinsically shallow depth of field (DoF) of conventional microscopes. Here, we propose an ultrafast, wide-field pump–probe interferometric microscope on a telecentric platform that significantly extends the effective DoF to ~18 μm at a high NA of 0.9 while maintaining high spatial resolution (down to 235 nm) and temporal resolution (~170 fs). The system enables single-frame 3D topography reconstruction without axial scanning or multi-view acquisition. We demonstrate these capabilities by capturing axial material flow during laser-induced microsphere melting that remain unobservable with conventional narrow-DoF systems, and by tracking the azimuthal rotation of ablation lobes during axial propagation of temporal focused spatiotemporal optical vortex (TF-STOV) pulses, directly revealing the spatiotemporal evolution of STOV-matter interactions.**

## 1. Introduction

Understanding ultrafast dynamic processes, such as femtosecond laser ablation, phase transitions, carrier transport, and structured-light-matter interactions, increasingly demands three-dimensional (3D) spatiotemporal characterization[1-7], because many of these phenomena involve material redistribution, energy transport, or morphological evolution along the optical axis that cannot be captured by planar observation alone. Over the past two decades, pump-probe microscopy, in which a pump pulse triggers the dynamics of interest and a time-delayed probe pulse freezes the transient state into a single wide-field snapshot, has established itself as the primary tool for accessing femtosecond-to-nanosecond dynamics[1-4]. Despite this capability, most pump-probe implementations remain confined to two-dimensional (*x*-*y*) observation at each time step, leaving axial information unresolved. The fundamental reason is an inherent trade-off in optical microscopy: high numerical aperture (NA) objectives, essential for fine lateral resolution, possess a depth of field (DoF) that scales as $\lambda/NA^2$ and is therefore restricted to roughly a few hundred nanometers to one micrometer at high NA.

One natural strategy to recover axial information is mechanical *z*-scanning. In wide-field imaging modalities, components such as tunable lens [8], acoustic lens [9], or deformable mirror [10] can be employed to rapidly scan the imaged plane along the optical axis. However, in the pump-probe context, *z*-scanning requires acquiring the complete time-delay series at each axial position, thereby multiplying the total acquisition time by the number of focal planes. Beyond *z*-scanning, point spread function (PSF) engineering, such as Bessel beam excitation[11-13], extends the excitation PSF but still requires lateral raster scanning that is incompatible with wide-field ultrafast imaging. Alternative techniques such as deep-learning-assisted holographic methods [14] enable single-frame extended-DoF imaging but depend on computational phase recovery that may introduce artifacts. Light field microscopy [15,16] captures angular information for single-frame volumetric reconstruction but at a fundamental cost to spatial resolution. In our previous work, we developed a single-pulse structured-light microscopy (SPSLM) technique [17] that projects structured illumination onto the sample surface through the microscope objective, achieving single-frame 3D profilometry within a pump-probe configuration. However, because the structured illumination is formed by interference of two beams passing through the objective, extending the axial fringe range requires reducing the crossing angle, which in turn enlarges the fringe period and degrades the lateral measurement resolution. This imposes a fundamental trade-off that limits the effective DoF to approximately 1.9 μm at NA 0.9. While individually powerful, none of these approaches simultaneously deliver quantitative 3D information with high-NA lateral resolution, femtosecond temporal resolution, and extended DoF.

In this work, we propose an ultrafast wide-field imaging system that circumvents the intrinsic DoF limitation of high-NA pump-probe microscopy by combining telecentric illumination with femtosecond interferometry. In this approach, the achievable DoF is determined by the temporal coherence length of the probe pulse rather than the geometric focal depth of the objective. Operating at a high NA of 0.9, the system achieves a temporal resolution of 170 fs, a spatial resolution of down to 235 nm and an extended DoF of approximately 18 μm, significantly surpassing the performance of our previous ultrafast 3D imaging technology, SPSLM[17]. Furthermore, we demonstrate the system's utility to capture ultrafast laser-induced melting of micrometer-sized microspheres, where the melt front progresses to depths of ~8 μm, far exceeding the conventional DoF. We also demonstrate the system's capability to track the axial evolution of light-matter interactions mediated by spatiotemporal optical vortex (STOV), highlighting its application for quantitative studies of topological light-matter interactions. Our results establish a practical route to high-NA, femtosecond-resolved, extended-DoF 3D imaging, filling a key gap between ultrafast temporal access and large-depth-range 3D imaging.

## 2. System and principle

The proposed experimental system is illustrated in Fig. 1a. A femtosecond laser beam, centered at 800 nm wavelength with a repetition rate of 10 Hz, pulse width of 120 fs, and single pulse energy of 6 mJ, is produced by an amplified Ti: sapphire laser system (Coherent, Legend Elite HE). This beam is split into two paths by using a dichroic mirror (DM1): one serves as an 800 nm-wavelength pump beam (red line) to induce ablation structures, while the other path is converted into a 400 nm-wavelength probe beam (violet line) via a BBO crystal for in-situ measurement of the fabricated structures.

For the pump beam, a single pulse is selected using a high-speed electromechanical shutter. The pump pulse power is adjusted by a half-wave plate (λ/2) and a polarizer (P1). The pump and probe beams are combined and directed into a microscope objective (OB2) using a dichroic mirror (DM2). A short-pass filter (FL) placed before the CCD transmits only the 400 nm probe light, blocking scattered pump light. The temporal delay between the pump and probe pulses is precisely controlled by adjusting a delay line in the probe beam path. By capturing images at different delay times, time-resolved images of the sample's transient response are obtained.

The core idea of this work is to make the DoF governed by the temporal coherence length of the probe pulse rather than the geometric focal depth of the objective. To realize this, the probe beam is implemented within a dual-telecentric interferometric configuration, as shown in Fig. 1b. The probe beam is split by BS1 into a sample arm and a reference arm. In the sample arm, the beam is focused by lens L5 onto the back focal plane (BFP) of objective OB2, emerging as a collimated wave that illuminates the sample at normal incidence. On the detection side, the tube lens (TL) is placed at its focal distance from the exit pupil of OB2, making the system telecentric in both object and image space. This geometry guarantees constant magnification independent of the sample's axial position. More fundamentally, because the probe beam illuminates the sample as a collimated wave at normal incidence, the optical path for light reflected from a surface element at height h is simply $2h$, yielding a phase shift $\Delta\varphi=4\pi h/\lambda$ that is strictly proportional to $h$. Crucially, this proportionality is preserved at all axial positions within the extended depth range, since the normal-incidence condition is independent of defocus. In the reference arm, the beam follows a deliberately symmetric layout: L5 focuses the light onto the BFP of a second, nominally identical objective (OB1), producing a collimated beam that illuminates a flat reference mirror. The reflected reference light is collected by OB1, relayed by TL, and directed onto the CCD.

Due to the broad spectral bandwidth inherent to femtosecond pulses, the probe pulse accumulates significant group-velocity dispersion (GVD) upon traversing the glass elements in the optical path. GVD introduces temporal chirp that both broadens the pulse duration and degrades the interference fringe contrast, thereby compromising temporal resolution and depth measurement fidelity. To overcome this issue, two strategies were adopted. First, both interferometer arms employ nominally identical objectives (OB1 and OB2) and symmetric optical path layouts, so that both pulses accumulate nearly identical GVD and wavefront aberrations, preserving high fringe contrast in broadband femtosecond interference. Second, the grating-pair compressor within the laser amplifier is fine-tuned to pre-compensate the accumulated GVD in the probe path, restoring the pulse duration to ~170 fs at the sample plane (**see Appendix B**).

Off-axis carrier fringes are generated by introducing a small angular tilt to the reference beam via mirror R5. The tilt angle sets the carrier fringe period, which is chosen to satisfy the Nyquist sampling condition of the CCD while providing sufficient spectral separation for frequency-domain phase extraction. Height information is obtained from the phase map extracted from each single-frame interferogram using Fourier Transform Profilometry (FTP) [18-20] through a linear calibration (**see Appendix A**). In this system, fringes persist as long as the sample and reference pulses overlap temporally at the detector. The depth at which this overlap vanishes is governed by the temporal coherence length of the probe pulse. It is in this sense that the DoF is transferred from a geometric quantity ($\sim\lambda/NA^2$) to a coherence quantity.

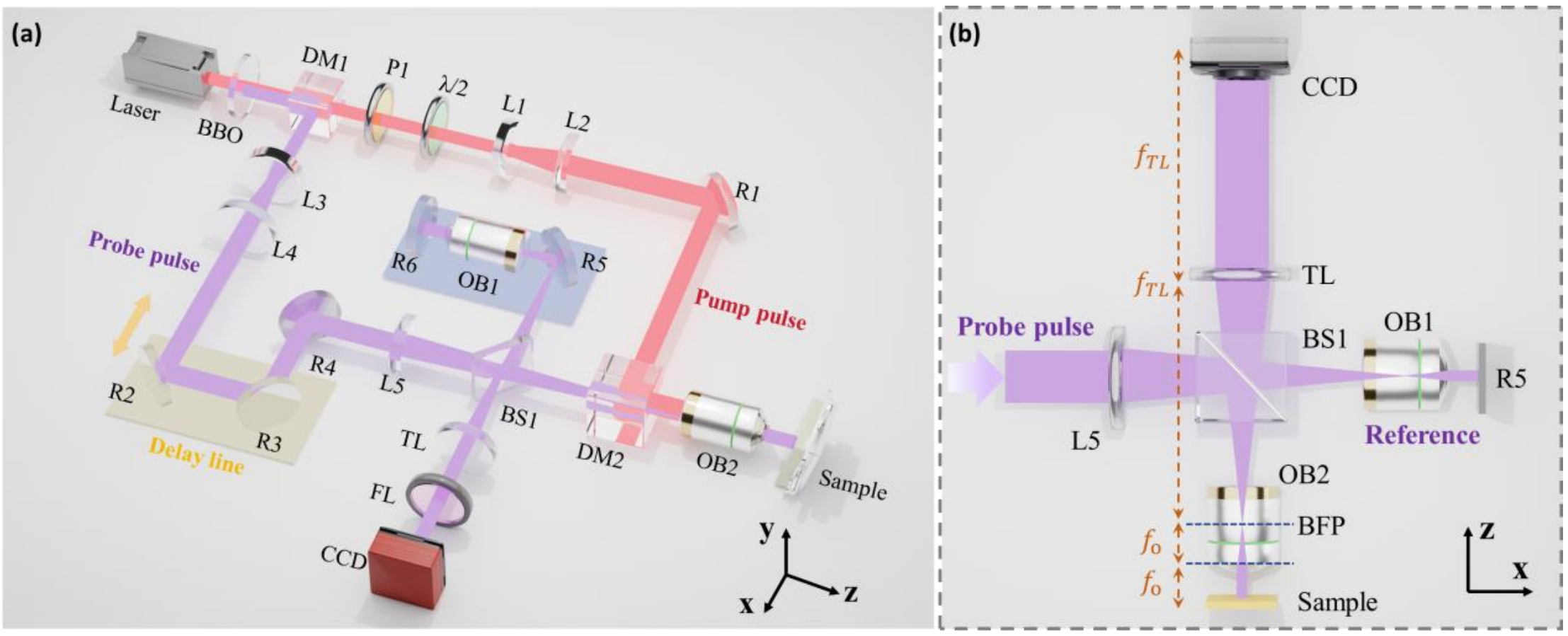

Fig. 1. (a) Schematic diagram of the ultrafast extended-DoF wide-field imaging system. A dichroic mirror splits the beam after BBO into 800 nm pump light (red) and 400 nm probe light (violet), and multiple short-wave mirrors are used in the probe optical path to completely filter out the 800 nm light; (b) Enlarged imaging block diagram in the probe optical path. The probe light is split by BS1 into a sample arm (downward) and a reference arm (right), and then focused by lens (L5) onto the back focal planes (BFPs) of the objectives in both the sample arm and reference arm, forming collimated femtosecond beams.

## 3. Results and discussion

To validate the ultrafast extended-DoF imaging capability of our system, we compared the proposed system with our previously developed SPSLM technique[17]. In the SPSLM approach, as depicted in Fig. 2a1, the probe beam is modulated by a digital micromirror device (DMD) loaded with stripe phase, generating the +1st and 0th diffraction orders. The two diffraction orders are then focused by a lens onto the back focal plane (BFP) of the microscope objective, and form interference fringes at the sample surface, resulting in tilted structured-light illumination. In this case, the axial extent of the structured probe beam is mainly determined by the crossing angle and ultimately the effective NA of the objective lens. In the extended-DoF configuration (Fig. 2b1), the collimated telecentric illumination described in Section 2 produces interference fringes whose visibility extends over the full coherence length of the probe pulse.

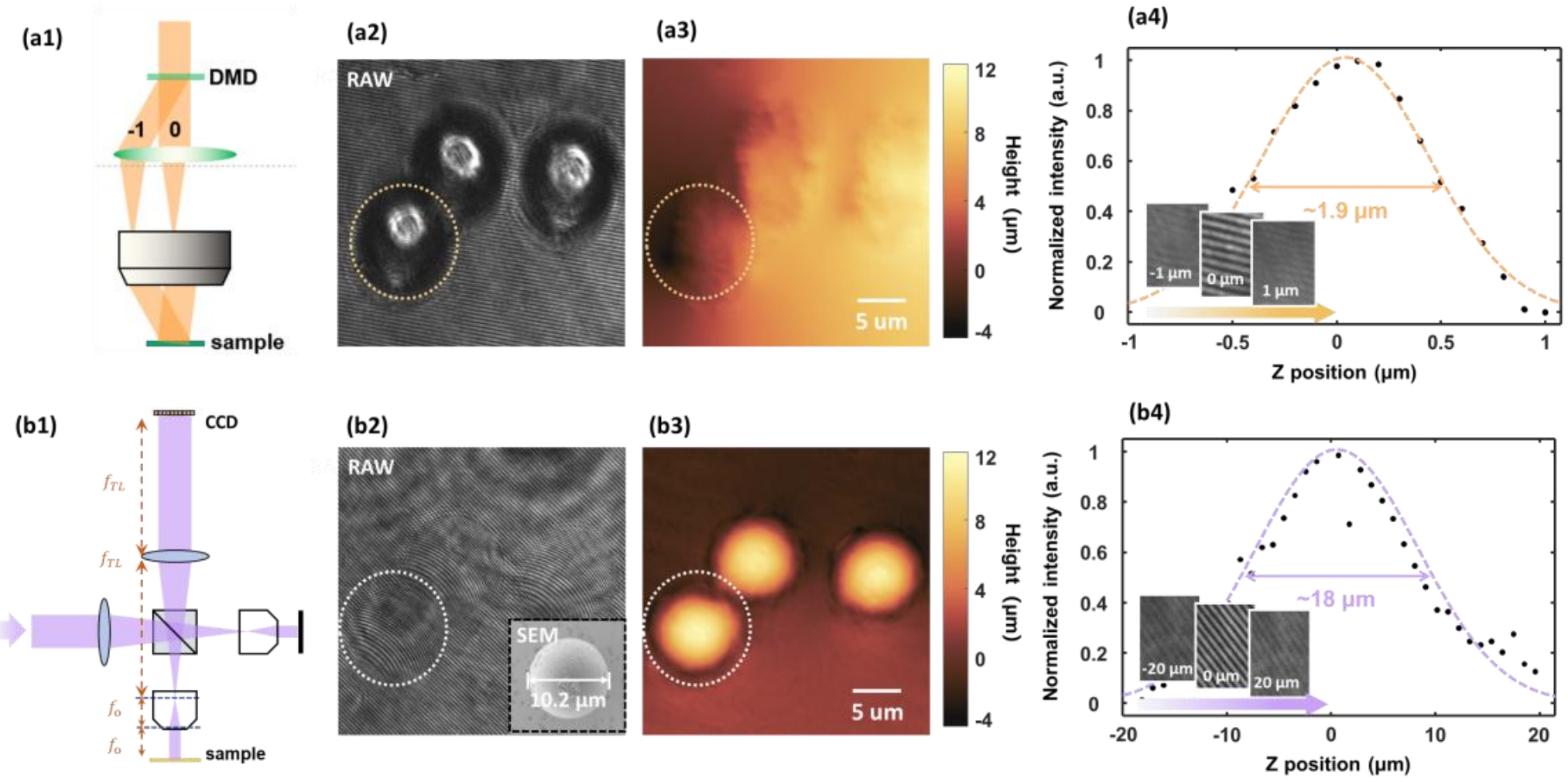

Fig. 2. DoF comparison between SPSLM (top) and extended-DoF wide-field imaging (bottom), quantified through axial fringe analysis and deep-structure microsphere imaging. (a1) Schematic of SPSLM imaging; (b1) Schematic of extended-DoF wide-field imaging. (a2, b2) Experimentally acquired raw fringe images demonstrating that SPSLM confines fringes within the objective's focal depth, while extended-DoF wide-field imaging achieves full coverage across the 10-μm-height PS microsphere (The black dashed box in (b2) contains the SEM image of PS spheres).(a3, b3) Reconstructed height profiles showing complete PS microsphere reconstruction via extended-DoF wide-field imaging (b3), contrasted with inaccurate reconstruction under SPSLM(a3). (a4, b4) The fringe visibility range at different axial positions measurements for both modalities, with Gaussian-fitted FWHM values of 1.9 μm (a4) and 18 μm (b4).

To directly compare the two modalities, we imaged transparent polystyrene (PS) spheres with a diameter of 10 μm diameter attached to a silicon (Si) substrate using both imaging modalities. In the SPSLM case (Fig. 2a2), interference fringes are visible only on the flat substrate surrounding the microsphere. Within the microsphere region, whose height far exceeds the ~1.9 μm DoF, no fringes are formed and the reconstructed height map (Fig. 2a3) fails to recover the sphere profile. In contrast, In the extended-DoF mode (Fig. 2b2), the collimated probe beam maintains coherent interference across the full depth of the microsphere, producing well-defined fringes over the entire surface.

The resulting height map (Fig. 2b3) accurately reproduces the complete sphere topography.

To quantify the DoF of each modality, we measured the normalized fringe visibility as a function of axial position. The Gaussian-fitted FWHM values are 1.9 μm for SPSLM (Fig. 2a4) and 18 μm for the extended-DoF system (Fig. 2b4), a nearly tenfold extension. This directly explains the microsphere imaging results: the 10 μm sphere height lies well within the 18 μm DoF of the extended system but far exceeds the 1.9 μm DoF of SPSLM.

We further characterized the lateral resolution of both modalities at different axial positions using a standard step sample with a height change of 83.5 ± 2.8 nm (Bruker DektakXT). The step sample was moved to various axial positions, and the corresponding step images were reconstructed for the SPSLM (Fig. 3a) and the extended-DoF imaging system (Fig. 3b). From the reconstructed images, we derivative the Line Spread Function (LSF) and subsequently its Fourier transformation, the Modulation Transfer Function (MTF). The spatial resolution was then calculated from the frequency at which the MTF value dropped to 3%, following the method described in Refs. [21-22]. The lateral resolution as a function of axial displacement was plotted in Fig. 3c and Fig. 3d, with the images clearly demonstrating the enhanced DoF. If we define a lateral resolution threshold of 500 nm as a benchmark, the SPSLM method exhibits a sharp decline in resolution beyond $|z| = \pm 0.7$ μm, resulting in blurred step images. Furthermore, due to the limitations of the objective pupil, reflected light from off-axis regions away from the focal plane (especially in the $+z$ direction) cannot be collected by the objective. Consequently, reliable resolution results cannot be obtained for $z > 0.4$ μm. In contrast, the extended-DoF wide-field imaging system maintains a lateral resolution below 500 nm even at axial positions of $|z| = \pm 9.6$ μm, presenting a nearly order-of-magnitude improvement in DoF. At the focal plane ($z = 0$ μm) in Fig. 3d, the spatial lateral resolution was measured to be 235 nm, closely aligning with the diffraction limit of the objective lens (~222 nm).

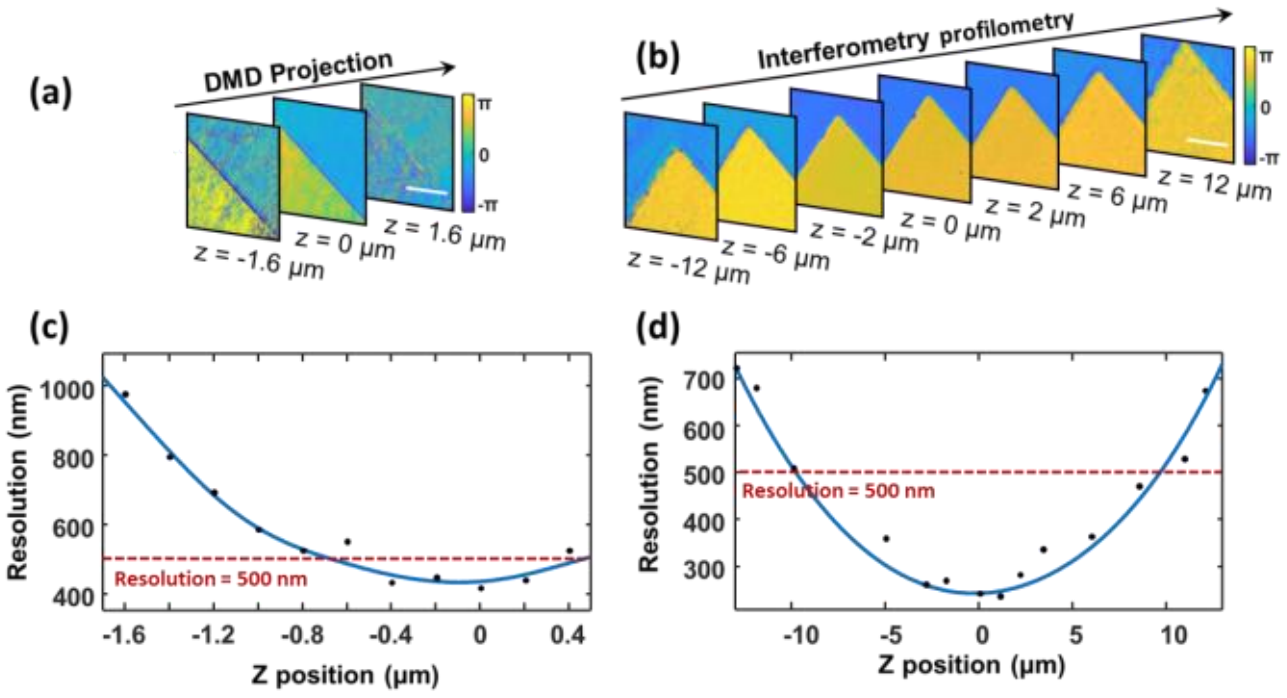

Fig. 3. Lateral resolution comparison at different axial positions for a standard step sample using (a) SPSLM and (b) extended-DoF wide-field imaging. Scalebar: 10 μm. (c) The SPSLM achieves step reconstruction (500 nm resolution) only within $|z| = \pm 0.7$ μm. (d) The extended-DoF system maintains sub-500 nm resolution within $|z| = \pm 9.6$ μm.

## 4. Extended-DoF wide-field spatiotemporal imaging of the ultrafast dynamics in laser ablation

### A. 3D ultrafast imaging of femtosecond-laser-induced melting dynamics in a microsphere

To demonstrate the ultrafast extended-DoF imaging capability of our system, we performed transient imaging of the melting dynamics in a polystyrene (PS) microsphere following femtosecond pulse excitation. The sample of 10-μm-diameter PS microsphere was sufficiently diluted in a liquid environment, then dripped onto a silicon wafer and allowed to settle for an extended period, ensuring deposition at the silicon (Si) substrate. As illustrated in Fig. 4a, the pump beam (Red beam, Fig. 4a) was precisely focused onto the apex of th e PS microsphere through the microscope objective, initiating melting within the liquid environment.

The energy of a single pump pulse was set to 4.8 nJ, with a beam radius of 2.34 μm, yielding a fluence of 28 mJ $cm^{-2}$. This value exceeds the thermal degradation threshold of polystyrene microspheres (10 mJ $cm^{-2}$) [23], thereby was sufficient to induce gradual melting process of the PS microsphere via thermal effects, enabling our system to track height variations. Fig. 4b presents ultrafast wide-field topography images and 3D renderings of the relative height change ($\Delta h$) of the PS microsphere at various time delays after pulse excitation. The relative height change $\Delta h$ was calculated as $\Delta h = h(0) - h(t)$, which is the difference between the height at time t and at initial height without ablation.

Due to femtosecond laser-induced thermal effects, a slight height change is observable at 100 ps (see Supplementary Visualization 1.avi). As shown in Fig. 4b, a depression crater forms centered at the laser focus (1.8 ns, depth ~2 μm). The crater depth progressively increases over time, reaching a melt depth of approximately 8 μm at 10.8 ns. These depth variations, extending beyond the conventional DoF of objective lens, are successfully resolved by our extended-DoF imaging system. The PS microsphere ultimately undergoes complete melting due to intense thermal effects, which process exceeds the temporal range covered by our delay line and thus is not shown here. Fig. 4c displays cross-sectional profiles at the arrow-marked positions (1.8 ns, 5.8 ns and 10.8 ns), and Fig. 4d shows the temporal evolution of the relative height change $\Delta h$ during PS microsphere melting, indicating a predominantly linear melting rate.

Furthermore, a series of pressure waves generated by the liquid-phase ablation process were captured (**see Appendix C**). Fig. 4e shows the extracted radii of the pressure waves at different time points, where the data points represent experimental results obtained from the images in the dashed box, and the curve shows the corresponding fitting results. The propagation speed of the pressure waves was derived from the derivative of the radius-time relationship and determined to be 1.54 km/s—closely matches the speed of sound in water (1.48 km/s, black line in Fig. 4e). Previous studies on underwater ablation dynamics [24] typically reported pressure wave speeds based on reflectivity changes. This result indicates that the propagation of pressure waves can also be quantitatively studied using our system.

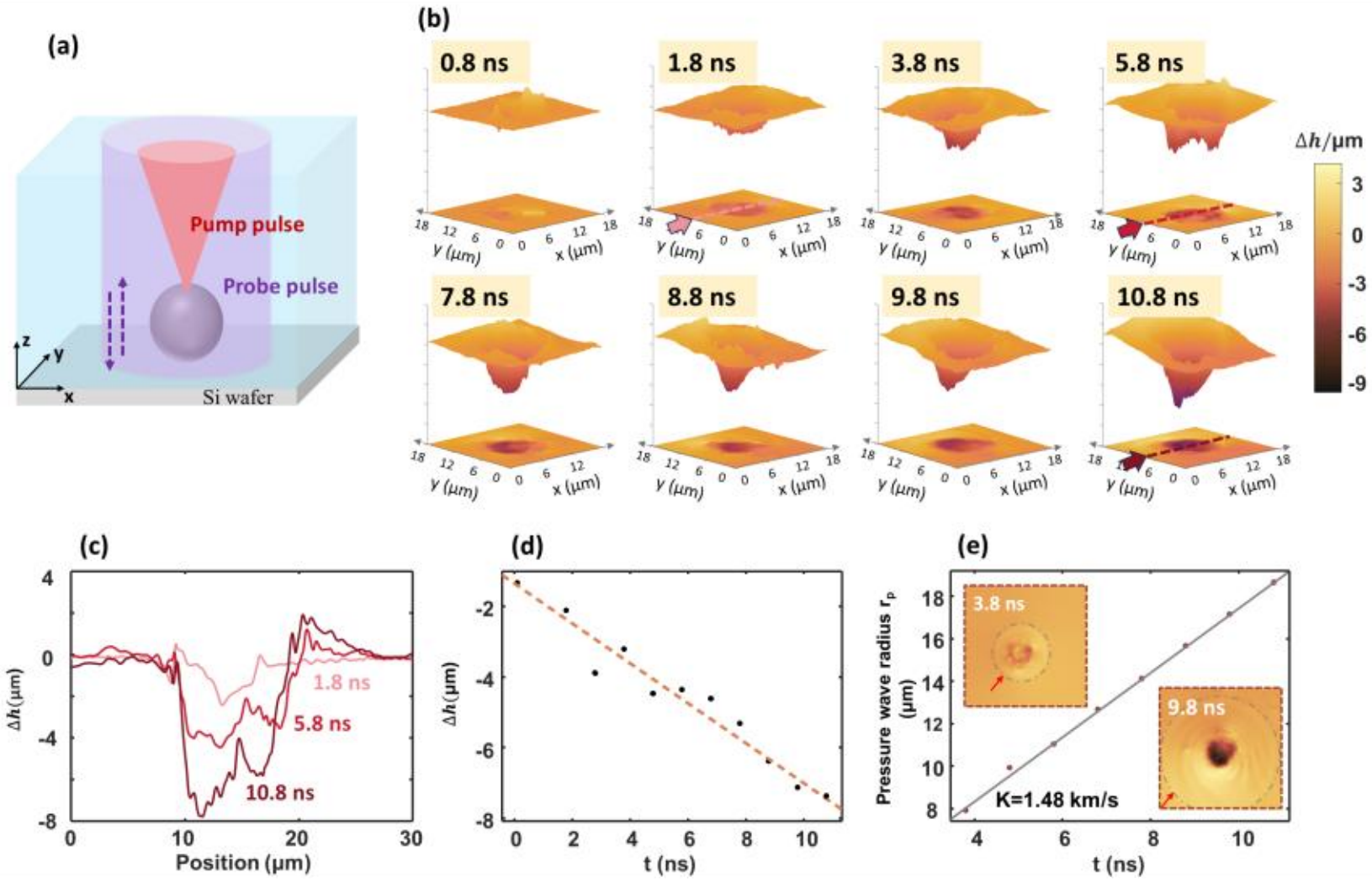

Fig. 4. 3D ultrafast extended-DoF imaging of femtosecond-pulse-induced melting dynamics in a PS microsphere. (a) Experimental schematic. (b) Time-resolved topography images of the relative height change ($\Delta h$) of PS microsphere at selected time delays post-excitation; The top images show 3D rendered views, and the bottom images display corresponding topography mapped onto the x-y plane, with height represented by a color scale. (c) Cross-sectional profiles at three arrow-marked positions (at 1.8 ns, 5.8 ns and 10.8 ns) in (b); (d) Temporal evolution of relative height change ($\Delta h$), revealing linear melting progression. The black dots represent experimental data points, obtained by averaging over a 5×5 pixel area at consistent cross-sectional positions; the orange dashed line denotes the fitted curve;(e) Propagation speed of the transient pressure wave induced by liquid-phase ablation in Appendix C. The insets show 2D relative height images at 3.8 ns and 9.8 ns, where the gray dashed circles mark the propagating wavefront positions of the pressure wave at the corresponding time instances.

### B. ultrafast dynamics of STOV-induced ablation

As a novel category of structured light carrying transverse orbital angular momentum (OAM), STOV exhibits a spiral phase distribution in the spatiotemporal domain, and has attracted considerable interest in the field of photonics [25-26]. Recently, we have demonstrated STOV with tunable transverse OAM and self-similar co-scaling in both time and space at off-focus positions using a temporal focusing apparatus [27]. Nonetheless, the STOV-material interactions, particularly the ultrafast dynamics of STOV-induced ablation, remain unexplored. Since STOV exhibit complex evolution during axial propagation, there is a critical need for in-situ, ultrafast wide-field imaging technology with extended-DoF to dynamically characterize both on-focus and off-focus STOV-material interactions. In Fig. 5, the ultrafast extended-DoF wide-field imaging system is employed to investigate STOV axial evolution and STOV-material interaction mechanisms. Compared to conventional spatiotemporal imaging modalities constrained by limited DoF, the ultrafast extended-DoF wide-field imaging system is better suited for characterizing the etching processes of ultrafast 3D light fields whose cross-sectional (*xy*-plane) morphology varies during axial propagation.

Fig. 5a illustrates the experimental pump beam modulation system. An adjustable slit (AS) is positioned within the pump optical path, while a pair of reflective blazed gratings (G1, G2; 1200 pl/mm) disperse the spatial frequency spectrum. The dispersed beam is subsequently directed onto a spatial light modulator (SLM) to impose spiral phase modulation for STOV generation. The generated STOV is focused onto the silicon sample surface through an objective lens, while the detection optical path remains identical to that shown in Fig. 1b. Fig. 5b shows the time-integrated spatial field distributions of STOVs in xy-planes at different axial positions obtained from theoretical calculations (upper) and experimental measurements (lower). The experimental results were acquired by axially changing the sample plane and capturing the corresponding transverse field distributions reflected to the CCD. A well-defined intensity singularity (null point) is observed in the center of STOV distribution. Notably, the intensity distribution of STOV displays rotational behavior around this singularity during axial propagation, with experimental results closely matching theoretical predictions.

Fig. 5c presents the transient evolution of silicon surface ablation morphology induced by STOV pulses at distinct axial propagation positions, obtained through axial movement of the sample stage. The spiral phase of the STOV produces non-uniform spatiotemporal intensity distribution, resulting in a distinct double-lobe ablation structure. This double-lobe structure undergoes azimuthal rotation as the axial position varies, well agreeing with the rotational behavior in Fig. 5b. Specifically, within the propagation range from $z$ = 0 μm to $z$ = 4 μm, the ablation lobes rotate by approximately 15°. The ultrafast extended-DoF wide-field imaging system enables direct acquisition of ablation profiles in the *xy*-plane at various axial positions without the need for realignment of the probe beam. Consequently, this approach offers a convenient and powerful tool for in-situ characterization and parameter optimization of ablation processes induced by complex spatiotemporal structured optical fields.

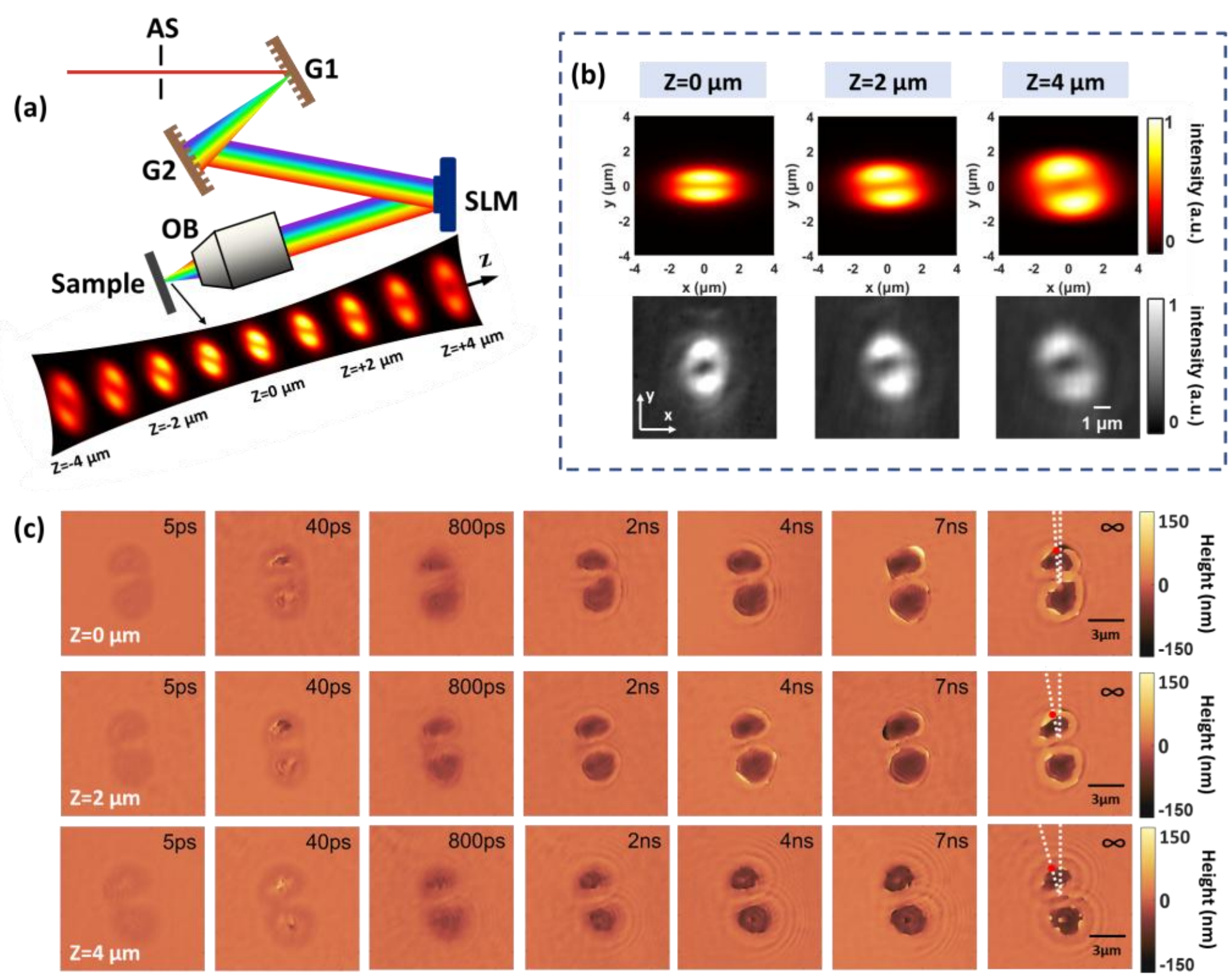

Fig. 5. Transient imaging results of STOV modulation and its interaction with materials. (a) Schematic diagram of STOV generation as the pump beam; (b) Light intensity distributions of generated STOV pulses in *xy*-planes at different axial positions obtained in theoretical calculations (upper) and experiments (lower), respectively; (c) Transient evolution process of ablation morphology on silicon surface induced by STOV pulses at different axial positions. Ablation profiles at each axial position were acquired by translating the sample stage, with the dual-telecentric configuration maintaining normal-incidence illumination and quantitative phase measurement fidelity throughout the extended DoF range.

## 5. Conclusions

In summary, this work presents an ultrafast extended-DoF wide-field imaging system that effectively addresses the inherent trade-off between spatial resolution and DoF in dynamic 3D imaging. By integrating pump-probe microscopy with dual-telecentric interferometry, the system achieves single-frame 3D topography reconstruction with a temporal resolution of 170 fs, spatial resolution down to 235 nm and an extended DoF of approximately 18 μm at NA 0.9, achieving an order of magnitude improvement compared to SPSLM. We demonstrate the system's capability through two distinct applications. First, in femtosecond-laser-induced melting of a polystyrene microsphere, the system resolves the progressive deepening of a depression crater with depth far exceeding the conventional geometric DoF, and reveals a predominantly linear melt front propagation rate over this temporal window. Second, in TF-STOV induced ablation of silicon, the system directly captures the azimuthal rotation of double-lobe ablation structures during axial propagation, in agreement with the theoretically prediction.

Looking forward, the present system opens several promising directions for future research and application. In the field of femtosecond laser micro/nanofabrication, the extended DoF enables real-time in-situ monitoring of high-aspect-ratio structures and axially varying ablation morphologies, as well as the investigation of ultrafast dynamics during fabrication processes, capabilities that have remained inaccessible to conventional narrow-DoF systems. In the field of structured light-matter interactions, the system's ability to simultaneously resolve axial structural evolution and femtosecond temporal dynamics provides a unique platform for characterizing complex three-dimensional light-matter interactions, particularly in scenarios where the transverse intensity and phase distributions undergo significant evolution along the propagation axis.

## APPENDIX A: Phase-to-height calibration.

A series of craters with varying depths were fabricated on a SiC surface by adjusting the fluence of femtosecond laser pulses. The machining beam was expanded using a beam expansion system comprising two lenses (fL1=125 mm, fL2 = -25 mm), resulting in an extended spot rather than a focused point on the sample surface. The precise topography of each crater was characterized using atomic force microscopy (AFM) (Fig. 6a), and the corresponding phase image was acquired via extended depth-of-field (DoF) wide-field imaging (Fig. 6b). Fig. 6c displays cross-sectional profiles of three craters with different depths, while Fig. 6d illustrates the relationship between the relative phase change derived from the phase images and the actual depth. A linear fit to this relationship established a quantitative phase-to-height conversion factor. Based on this calibration, the phase images reconstructed in experiments can be accurately mapped into height images.

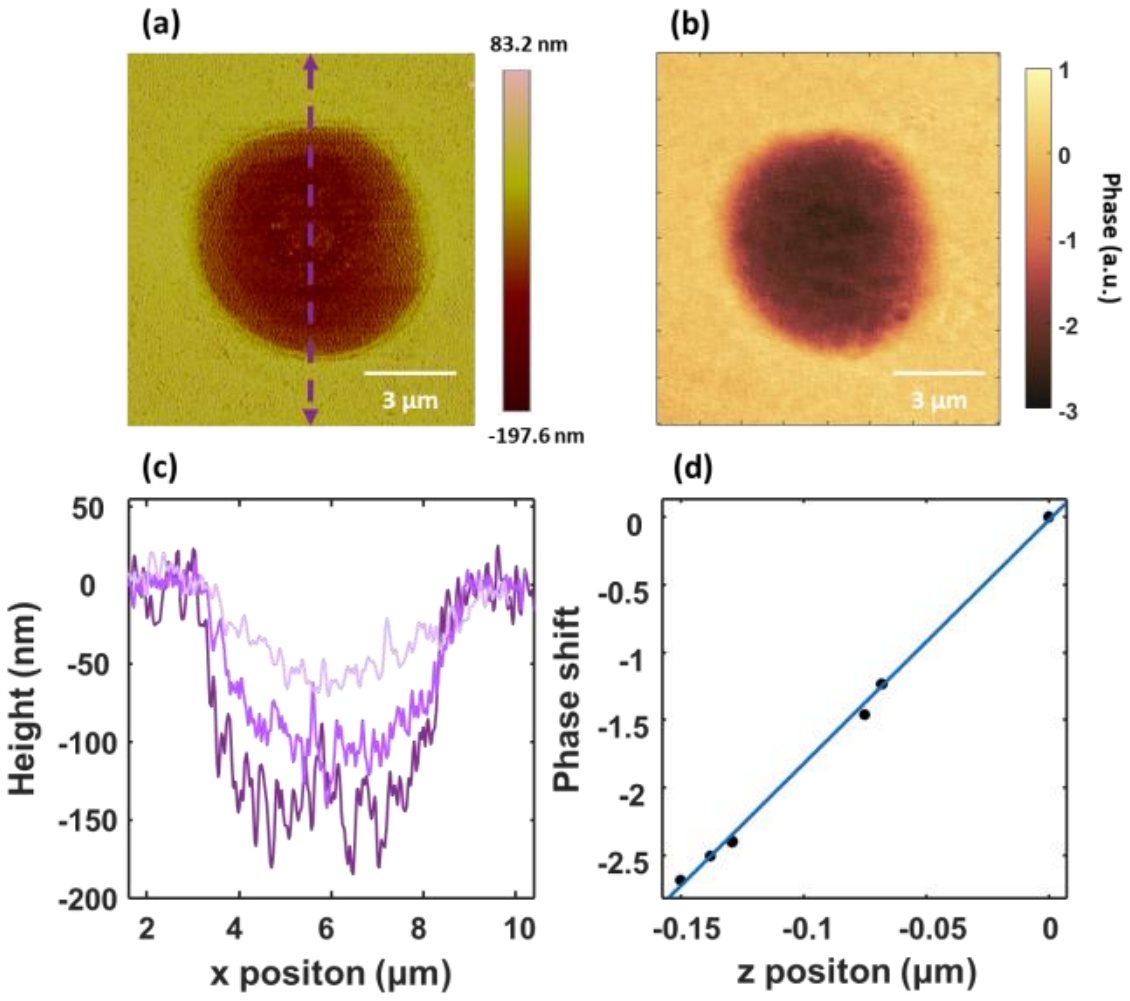

Fig. 6. Phase-to-height calibration of the experimental system. A series of craters with different depths were fabricated on a SiC surface via femtosecond single-pulse ablation with varying pulse energy levels. The same structures were characterized by atomic force microscopy (AFM) (a) and phase imaging (b) to obtain accurate height and phase profiles, respectively. (c) Cross-sectional height profiles taken along the central region (purple dashed line in (a)) of three representative craters with different depths. (d) Linear fitting of phase variation as a function of crater depth, confirming a well-defined phase-to-height proportionality.

## APPENDIX B: Calibration of temporal resolution.

For SPSLM, the series of dispersive elements in the optical path cause a broadening of the probe pulse width at the sample plane (compared to the 120 fs initial pulse). Based on theoretical calculations of the effective length and refractive index of each optical component at a wavelength of 400 nm, the probe pulse width at the sample plane was determined to be 256 fs [3,20], as shown in Fig. 7a.

In the extended-DoF wide-field imaging system, we employed a grating pair to compensate for the dispersion in the optical path and utilized an interferometric measurement approach to experimentally verify the improved temporal resolution. Experimentally, we adjusted the two interferometer arms of the probe light and measured the fringe modulation at different delays, which can be expressed as:

$$K = \frac{I_{max}-I_{min}}{I_{max}+I_{min}} \tag{1}$$

where K represents the fringe visibility, Imax represents the maximum light intensity, and Imin represents the minimum light intensity. We consider that the time profile of the broadened pulse is Gaussian, so the ratio of the pulse width to the fringe visibility waveform width is 2 . In this way, the probe pulse width after passing through the objective was measured to be 169.7 fs, as shown in Fig. 7b, where the black data points represent the normalized intensity values of the interference fringe modulation (K) obtained experimentally, and the purple dashed curve corresponds to the Gaussian fit.

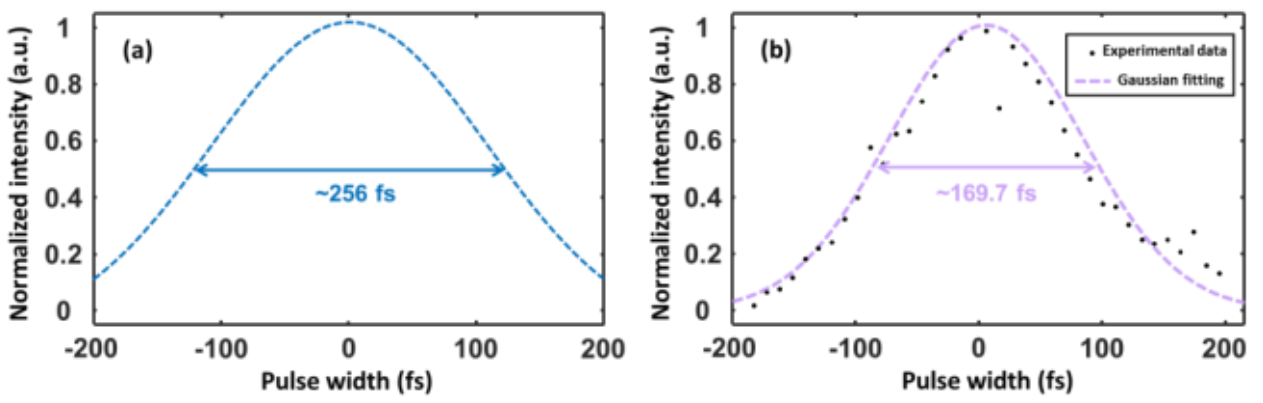

Fig. 7. Comparison of temporal resolution. (a) Theoretical calculation of the temporal resolution (256 fs) for SPSLM. (b) Measured pulse width after dispersion compensation. By adjusting the delay between the two interferometer arms of the probe pulse, we plotted the fringe modulation K curves at different delay positions and extracted the full width at half maximum (FWHM) of the time profile after Gaussian fitting.

## APPENDIX C: Ablation of PS microspheres in liquid.

In the experiment of laser-induced melting of polystyrene (PS) microspheres submerged in liquid, the relative height images were obtained by calculating the difference between the transient topographic images at various time delays after pulse excitation and the initial topography of the intact microsphere before laser arrival (t = 0 ps). This differential method enhances the visualization of transient height changes and minimizes artifacts caused by liquid environmental disturbances across repeated experiments. The reconstruction workflow is illustrated in Fig. 8a. Fig. 8b presents representative transient 2D topographic images of the microsphere. Furthermore, a series of pressure waves generated by the liquid-phase ablation process were captured.

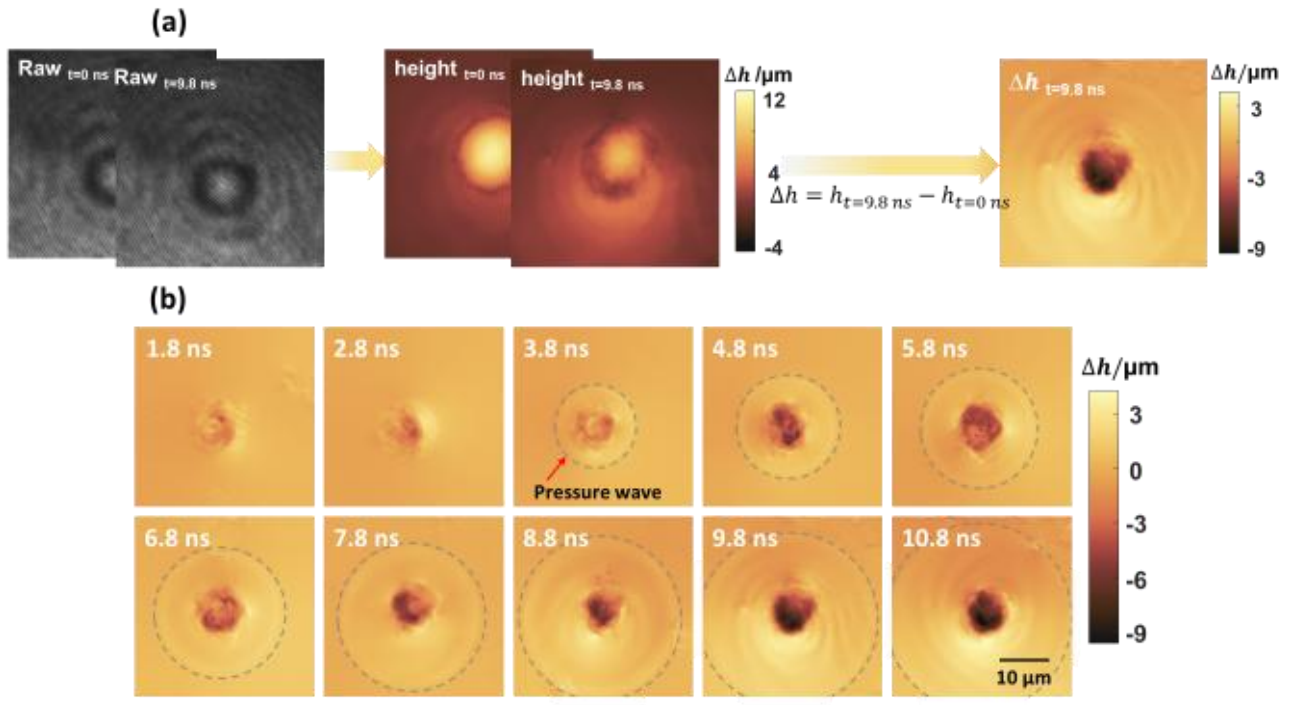

Fig. 8. (a) Workflow for processing height variation; (b) Transient 2D image of relative height change of the PS microsphere. The gray dashed circles in the figure indicate the wavefront positions of the pressure wave propagation at different time instances.

## Author contributions

Q. Wei, J. Ni, Y. Zhang and C. Min developed the concept presented in this work. Z. Tan, J. Pan, S. Zhang and Z. Zhou provided experimental support. Q. Wei performed the experiment data analysis and wrote the manuscript. J. Ni, C. Min revised the original draft. J. Ni, C. Min and X. Yuan supervised all the work, oversaw preparation of the manuscript, and acquired funding. All authors discussed the results and commented on the manuscript.

## Conflicts of interest

There are no conflicts to declare.

## Fundings

This research is supported by the National Natural Science Foundation of China ( 62575184, 62375177, and 92150301); Shenzhen Science and Technology Program (JCYJ20210324120403011, JCYJ20241202124428038, JCYJ20250604181103005, RCJC20210609103232046, RCJC20200714114435063); Natural Science Foundation of Guangdong Province (2026A1515010428) ; Research Team Cultivation Program of ShenZhen University (2023QNT014); Shenzhen University 2035 Initiative (2023B004); The authors acknowledge the Photonics Center of Shenzhen University for tec